\documentclass[journal]{IEEEtran}
\usepackage{cite}
\usepackage[pdftex]{graphicx}
\usepackage{amsmath}
\usepackage{amssymb}
\usepackage{array}
\usepackage{url}
\usepackage[misc,geometry]{ifsym}
\usepackage{booktabs}
\usepackage{multirow}
\usepackage{siunitx}
\usepackage{subcaption}
\usepackage[colorlinks=true]{hyperref}
\usepackage{url}
\usepackage{xcolor}
\usepackage[normalem]{ulem}

\begin{document}

\title{ADN: Artifact Disentanglement Network for \\ Unsupervised Metal Artifact Reduction}
%
%
%
\author{Haofu~Liao,~\IEEEmembership{Student Member,~IEEE,}
        Wei-An Lin,~\IEEEmembership{Student Member,~IEEE,}\\
        S. Kevin Zhou,~\IEEEmembership{Senior Member,~IEEE,}
        and~Jiebo~Luo,~\IEEEmembership{Fellow,~IEEE}
\thanks{H. Liao and J. Luo is with the Department of Computer Science, University of Rochester, Rochester, NY, USA (email: hliao6@cs.rochester.edu).}
\thanks{W. Lin is with the Department of Electrical and Computer Engineering, University of Maryland, College Park, MD, USA.}
\thanks{S. Zhou is with the Institute of Computing Technology, Chinese Academy of Sciences, Beijing, China and Peng Cheng Laboratory, Shenzhen, China.}}


\maketitle

\begin{abstract}
Current deep neural network based approaches to computed tomography (CT) metal artifact reduction (MAR) are supervised methods that rely on synthesized metal artifacts for training.  However, as synthesized data may not accurately simulate the underlying physical mechanisms of CT imaging, the supervised methods often generalize poorly to clinical applications. To address this problem, we propose, to the best of our knowledge, the first unsupervised learning approach to MAR. Specifically, we introduce a novel artifact disentanglement network that disentangles the metal artifacts from CT images in the latent space. It supports different forms of generations (artifact reduction, artifact transfer, and self-reconstruction, etc.) with specialized loss functions to obviate the need for supervision with synthesized data. Extensive experiments show that when applied to a synthesized dataset, our method addresses metal artifacts significantly better than the existing unsupervised models designed for natural image-to-image translation problems, and achieves comparable performance to existing supervised models for MAR. When applied to clinical datasets, our method demonstrates better generalization ability over the supervised models. The source code of this paper is publicly available at \url{https://github.com/liaohaofu/adn}.

\end{abstract}


%
\IEEEpeerreviewmaketitle

\section{Introduction}

\IEEEPARstart{M}{etal} artifact is one of the commonly encountered problems in computed tomography (CT). It arises when a patient carries metallic implants, e.g., dental fillings and hip prostheses. Compared to body tissues, metallic materials attenuate X-rays significantly and non-uniformly over the spectrum, leading to inconsistent X-ray projections. The mismatched projections will introduce severe streaking and shading artifacts in the reconstructed CT images, which significantly degrade the image quality and compromise the medical image analysis as well as the subsequent healthcare delivery. 

To reduce the metal artifacts, many efforts have been made over the past decades~\cite{mar_review}. Conventional approaches~\cite{li,nmar} address the metal artifacts by projection completion, where the metal traces in the X-ray projections are replaced by estimated values. After the projection completion, the estimated values need to be consistent with the imaging content and the underlying projection geometry. When the metallic implant is large, it is challenging to satisfy these requirements and thus secondary artifacts are often introduced due to an imperfect completion. Moreover, the X-ray projection data, as well as the associated reconstruction algorithms, are often held out by the manufactures, which limits the applicability of the projection based approaches.

\begin{figure}[t]
\centering
\includegraphics[width=0.8\linewidth]{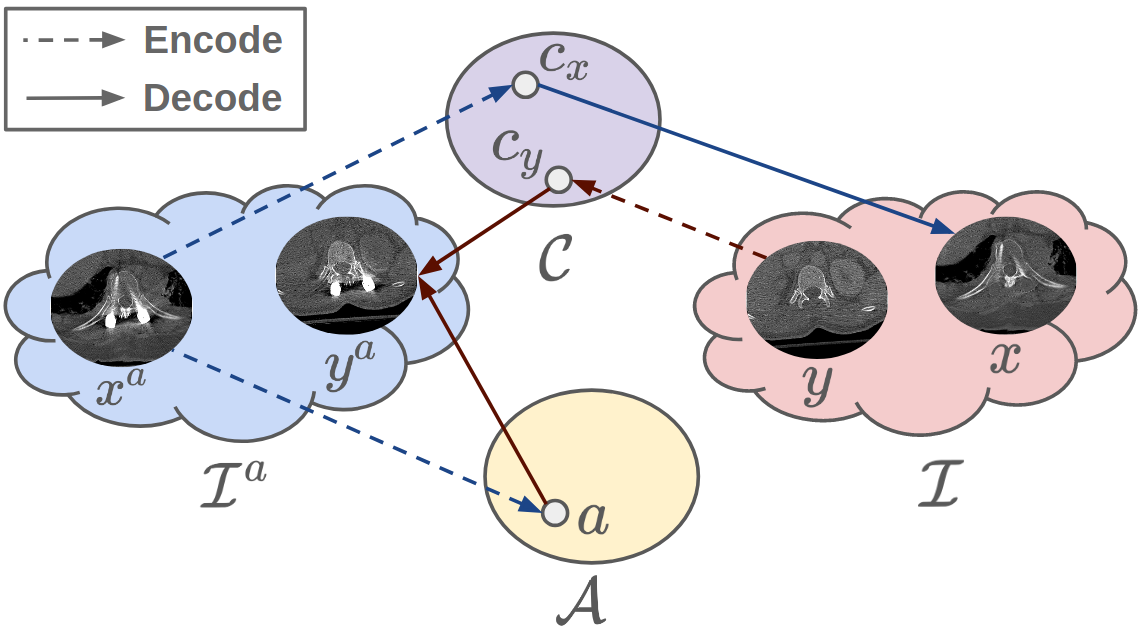}
\caption{Artifact disentanglement. The content and artifact components of an image $x^a$ from the artifact-affected domain $\mathcal{I}^a$ is mapped separately to the content space $\mathcal{C}$ and the artifact space $\mathcal{A}$, i.e., artifact disentanglement. An image $y$ from the artifact-free domain $\mathcal{I}$ contains no artifact and thus is mapped only to the content space. Decoding without artifact code removes the artifact from an artifact-affected image (blue arrows $x^a \rightarrow x$) while decoding with the artifact code adds artifacts to an artifact-free image (red arrows $y \rightarrow y^a$). }
\label{fig:assumption}
\end{figure}

A workaround to the limitations of the projection based approaches is to address the metal artifacts directly in the CT images. However, since the formation of metal artifacts involves complicated mechanisms such as beam hardening, scatter, noise, and the non-linear partial volume effect~\cite{mar_review}, it is very challenging to model and reduce metal artifacts in the CT images with traditional approaches. Therefore, recent approaches~\cite{cnnmartmi,cgan-mar,rl-arcnn,destreaknet} to metal artifact reduction (MAR) propose to use deep neural networks (DNNs) to inherently address the modeling of metal artifacts, and their experimental results show promising MAR performances. All the existing DNN-based approaches are supervised methods that require pairs of anatomically identical CT images, one with and the other without metal artifacts, for training. As it is clinically impractical to acquire such pairs of images, most of the supervised methods resort to synthesizing metal artifacts in CT images to simulate the pairs. However, due to the complexity of metal artifacts and the variations of CT devices, the synthesized artifacts may not accurately reproduce the real clinical scenarios, and the performances of these supervised methods tend to degrade in clinical applications.

In this work, we aim to address the challenging yet more practical unsupervised setting where {\it no paired CT images are available and required for training.} To this end, we reformulate the artifact reduction problem as an artifact disentanglement problem. As illustrated in Fig.~\ref{fig:assumption}, we assume that any artifact-affected image consists of an artifact component (i.e., metal artifacts, noises, etc.) and a content component (i.e., the anatomical structure). Our goal is to disentangle these two components in the latent space, and artifact reduction can be readily achieved by reconstructing CT images without the artifact component. Fundamentally, this artifact disentanglement without paired images is made possible by grouping the CT images into two groups, one with metal artifacts and the other without metal artifacts. In this way, we introduce an \textit{inductive bias}~\cite{disentanglement} that a model may inherently learn artifact disentanglement by comparing between these two groups. More importantly, the artifact disentanglement assumption guides manipulations in the latent space. This can be leveraged to include additional inductive biases that apply self-supervisions between the outputs of the model (See Sec.~\ref{sec:learn}) and thus obviate the need for paired images.

Specifically, we propose an artifact disentanglement network (ADN) with specialized encoders and decoders that handle the encoding and decoding of the artifact and content components separately for the unpaired inputs. Different combinations of the encoders and decoders support different forms of image translations (See Sec.~\ref{sec:enc_dec}), e.g., artifact reduction, artifact synthesis, self-reconstruction, and so on. ADN exploits the relationships between the image translations for unsupervised learning. Extensive experiments show that our method achieves comparable performance to the existing supervised methods on a synthesized dataset. When applied to clinical datasets, all the supervised methods do not generalize well due to a significant domain shift, whereas ADN delivers consistent MAR performance and significantly outperforms the compared supervised methods.

\section{Related work}

\textbf{Conventional Metal Artifact Reduction.} Most conventional approaches address metal artifacts in X-ray projections. A straightforward way is to directly correct the X-ray measurement of the metallic implants by modeling the underlying physical effects such as beam hardening~\cite{beam_harden1,beam_harden2}, scatter~\cite{scatter}, and so on. However, the metal traces in projections are often corrupted. Thus, instead of projection correction, a more common approach is to replace the corrupted region with estimated values. Early approaches~\cite{li2,li} fill the corrupted regions by linear interpolation which often introduces new artifacts due to the inaccuracy of the interpolated values. To address this issue, a state-of-the-art approach~\cite{nmar} introduces a prior image to normalize the X-ray projections before the interpolation.

\textbf{Deep Metal Artifact Reduction.} A number of studies have recently been proposed to address MAR with DNNs. RL-ARCNN~\cite{rl-arcnn} introduces residual learning into a deep convolutional neural network (CNN) and achieves better MAR performance than standard CNN. DestreakNet~\cite{destreaknet} proposes a two-streams approach that can take a pair of NMAR~\cite{nmar} and detail images as the input to jointly reduce metal artifacts. CNNMAR~\cite{cnnmartmi} uses CNN to generate prior images in the CT image domain to help the correction in the projection domain. Both DestreakNet and CNNMAR show significant improvements over the existing non-DNN based methods on synthesized datasets. Recently, adversarial learning has been widely used for MAR and the related artifact reduction. cGANMAR~\cite{cgan-mar} adapts Pix2pix~\cite{pix2pix} to further improve the DNN-based MAR performance. SVAR-GAN~\cite{svar_gan} introduces perceptual network~\cite{johnson2016perceptual} and a focused GAN loss to improve the adversarial learning. MPN~\cite{mpn} proposes to address MAR in the sinogram domain and converts the MAR problem to an image inpainting problem.  The authors introduces a mask pyramid network together with the adversarial learning for better inpainting performance. More recently, DuDoNet~\cite{dudonet} proposes to address MAR jointly in both the sinogram domain and the image domain. To achieve this goal, DuDoNet introduces a differentiable radon inversion layer that bridges the two domains for end-to-end learning.

\textbf{Unsupervised Image-to-Image Translation.} Image artifact reduction can be regarded as a form of image-to-image translation. One of the earliest unsupervised methods in this category is CycleGAN~\cite{cyclegancvpr} where a cycle-consistency design is proposed for unsupervised learning. MUNIT~\cite{muniteccv} and DRIT~\cite{driteccv} improve CycleGAN for diverse and multimodal image generation. However, these unsupervised methods aim at image synthesis and do not have suitable components for artifact reduction. Another recent work that is specialized for artifact reduction is deep image prior (DIP)~\cite{dipcvpr}, which, however, only works for less structured artifacts such as additive noise or compression artifacts.

\textbf{Preliminary work.} A preliminary version~\cite{adn} of this manuscript was previously published. This paper extends the preliminary version substantially with the following improvements.
\begin{itemize}
    \item We include more details (with illustrations) about the motivations and assumptions of the artifact disentanglement to help the readers better understand this work at high-level.
    \item We include improved notations and problem formulation to describe this work more precisely.
    \item We redraw the diagram of the overall architecture and add new diagrams as well as the descriptions about the detailed architectures of the subnetworks.
    \item We discuss the reasoning about the design choices of the loss functions and the network architectures to better inform and enlighten the readers about our work.
    \item We add several experiments to better demonstrate the effectiveness of the proposed approach. Specifically, we add comparisons with conventional approaches, add comparisons with different variants of the proposed approach for an ablation study, and add evaluations about the proposed approach on artifact transfer.
    \item We include discussions about the significance and potential applications of this work.
\end{itemize}

\section{Methodology}

\begin{figure}[t]
\centering
\includegraphics[width=\linewidth]{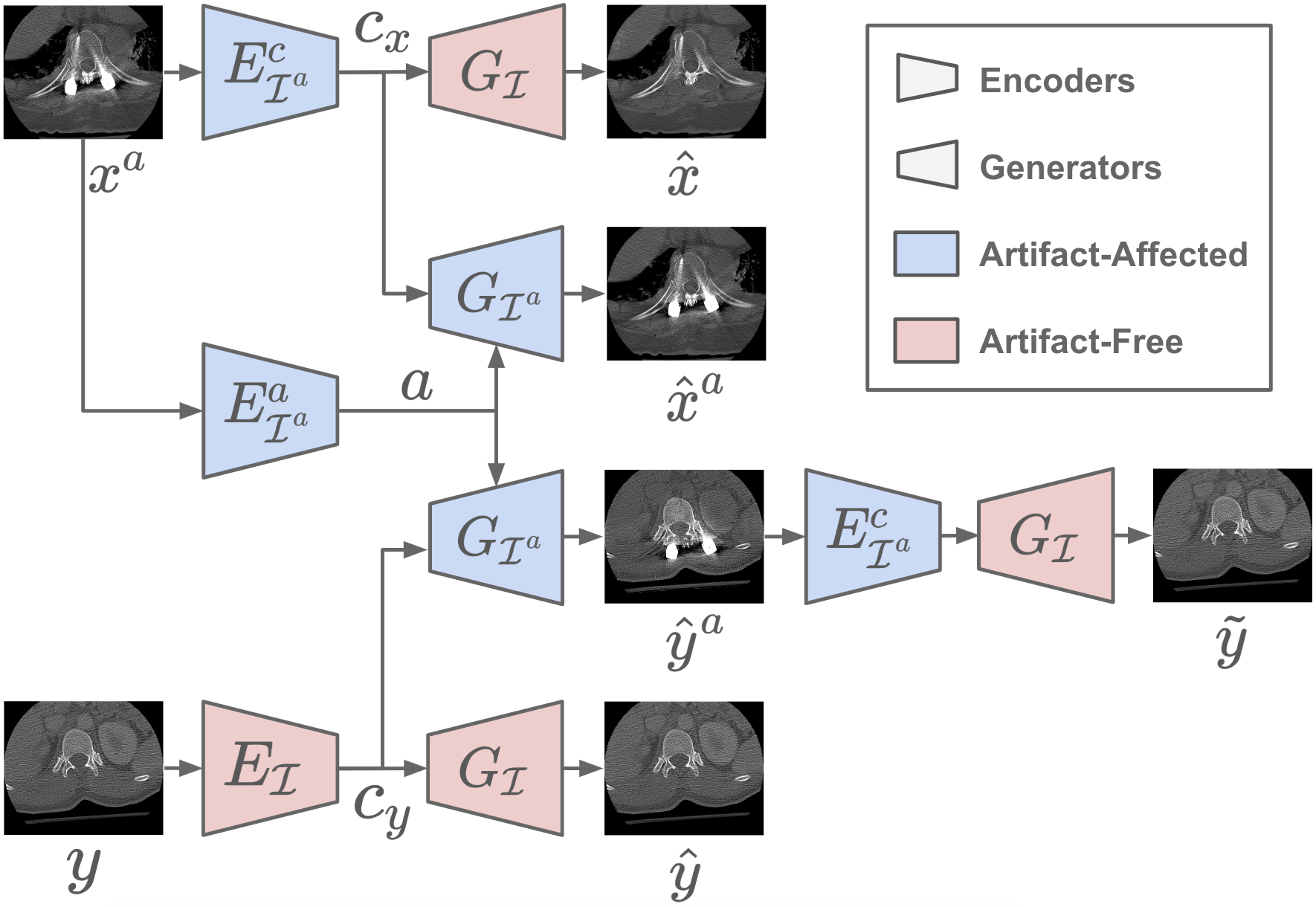}
\caption{Overview of the proposed artifact disentanglement network (ADN). Taking any two unpaired images, one from $\mathcal{I}^a$ and the other from $\mathcal{I}$, as the inputs, ADN supports four different forms of image translations: $\mathcal{I}^a \rightarrow \mathcal{I}$, $\mathcal{I} \rightarrow \mathcal{I}^a$, $\mathcal{I} \rightarrow \mathcal{I}$ and $\mathcal{I}^a \rightarrow \mathcal{I}^a$.  }
\label{fig:overview}
\end{figure}

Let $\mathcal{I}^{a}$ be the domain of all artifact-affected CT images and $\mathcal{I}$ be the domain of all artifact-free CT images. We denote $\mathcal{P} = \{(x^a, x) \mid x^a \in \mathcal{I}^{a}, x \in \mathcal{I}, f(x^a) = x\}$ as a set of paired images, where $f: \mathcal{I}^{a} \rightarrow \mathcal{I}$ is an MAR model that removes the metal artifacts from $x$. In this work, we assume no such paired dataset is available and we propose to learn $f$ with unpaired images.

As illustrated in Fig.~\ref{fig:assumption}, the proposed method disentangles the artifact and content components of an artifact-affected image $x^a$ by encoding them separately into a content space $\mathcal{C}$ and an artifact space $\mathcal{A}$. If the disentanglement is well addressed, the encoded content component $c_x \in \mathcal{C}$ should contain no information about the artifact while preserving all the content information. Thus, decoding from $c_x$ should give an artifact-free image $x$ which is the artifact-removed counterpart of $x^a$. On the other hand, it is also possible to encode an artifact-free image $y$ into the content space which gives a content code $c_y$. If $c_y$ is decoded together with an artifact code $a \in \mathcal{A}$, we obtain an artifact-affected image $y^a$. In the following sections, we introduce an artifact disentanglement network (ADN) that learns these encodings and decodings without paired data.

\subsection{Encoders and Decoders} \label{sec:enc_dec}

The architecture of ADN is shown in Fig.~\ref{fig:overview}. It contains a pair of artifact-free image encoder $E_{\mathcal{I}}: \mathcal{I} \rightarrow \mathcal{C}$ and decoder $G_{\mathcal{I}}: \mathcal{C} \rightarrow \mathcal{I}$ and a pair of artifact-affected image encoder $E_{\mathcal{I}^a} = \{E_{\mathcal{I}^{a}}^c: \mathcal{I}^a \rightarrow \mathcal{C}, E_{\mathcal{I}^{a}}^a: \mathcal{I}^a \rightarrow \mathcal{A}\}$ and decoder $G_{\mathcal{I}^a}: \mathcal{C} \times \mathcal{A} \rightarrow \mathcal{I}^a$. The encoders map an image sample from the image domain to the latent space and the decoders map a latent code from the latent space back to the image domain. Note that unlike a conventional encoder, $E_{\mathcal{I}^{a}}$ consists of a content encoder $E_{\mathcal{I}^{a}}^c$ and an artifact encoder $E_{\mathcal{I}^{a}}^a$, which encode the content and artifacts separately to achieve artifact disentanglement.

Specifically, given two unpaired images $x^a \in \mathcal{I}^{a}$ and $y \in \mathcal{I}$, $E_{\mathcal{I}^{a}}^c$ and $E_{\mathcal{I}}$ map the content component of $x^a$ and $y$ to the content space $\mathcal{C}$, respectively. $E_{\mathcal{I}^{a}}^a$ maps the artifact component of $x^a$ to the artifact space $\mathcal{A}$. We denote the corresponding latent codes as
\begin{equation}
     c_x = E_{\mathcal{I}^a}^c(x^a), a = E_{\mathcal{I}^{a}}^a(x^a),  c_y = E_{\mathcal{I}}(y).
\end{equation}
$G_{\mathcal{I}^{a}}$ takes a content code and an artifact code as the input and outputs an artifact-affected image.  Decoding from $c_x$ and $a$ should reconstruct $x^a$ and decoding from $c_y$ and $a$ should add artifacts to $y$,
\begin{equation} \label{eqn:output1}
\hat{x}^a = G_{\mathcal{I}^a}(c_x, a), \quad \hat{y}^a = G_{\mathcal{I}^a}(c_y, a)
\end{equation}
$G_{\mathcal{I}}$ takes a content code as the input and outputs an artifact-free image. Decoding from $c_x$ should remove the artifacts from $x^a$ and decoding from $c_y$ should reconstruct $y$,
\begin{equation} \label{eqn:output2}
\hat{x} = G_{\mathcal{I}}(c_x), \quad \hat{y} = G_{\mathcal{I}}(c_y).
\end{equation}
Note that $\hat{y}^a$ can be regarded as a synthesized artifact-affected image whose artifacts come from $x^a$ and content comes from $y$. Thus, by reapplying $E_{\mathcal{I}^{a}}^c$ and $G_{\mathcal{I}}$, it should remove the synthesized artifacts and recover $y$,
\begin{equation} \label{eqn:output3}
\tilde{y} = G_{\mathcal{I}}(E_{\mathcal{I}^{a}}^c(\hat{y}^a)).
\end{equation}

\subsection{Learning} \label{sec:learn}

\begin{figure}[t]
\centering
\includegraphics[width=0.95\linewidth]{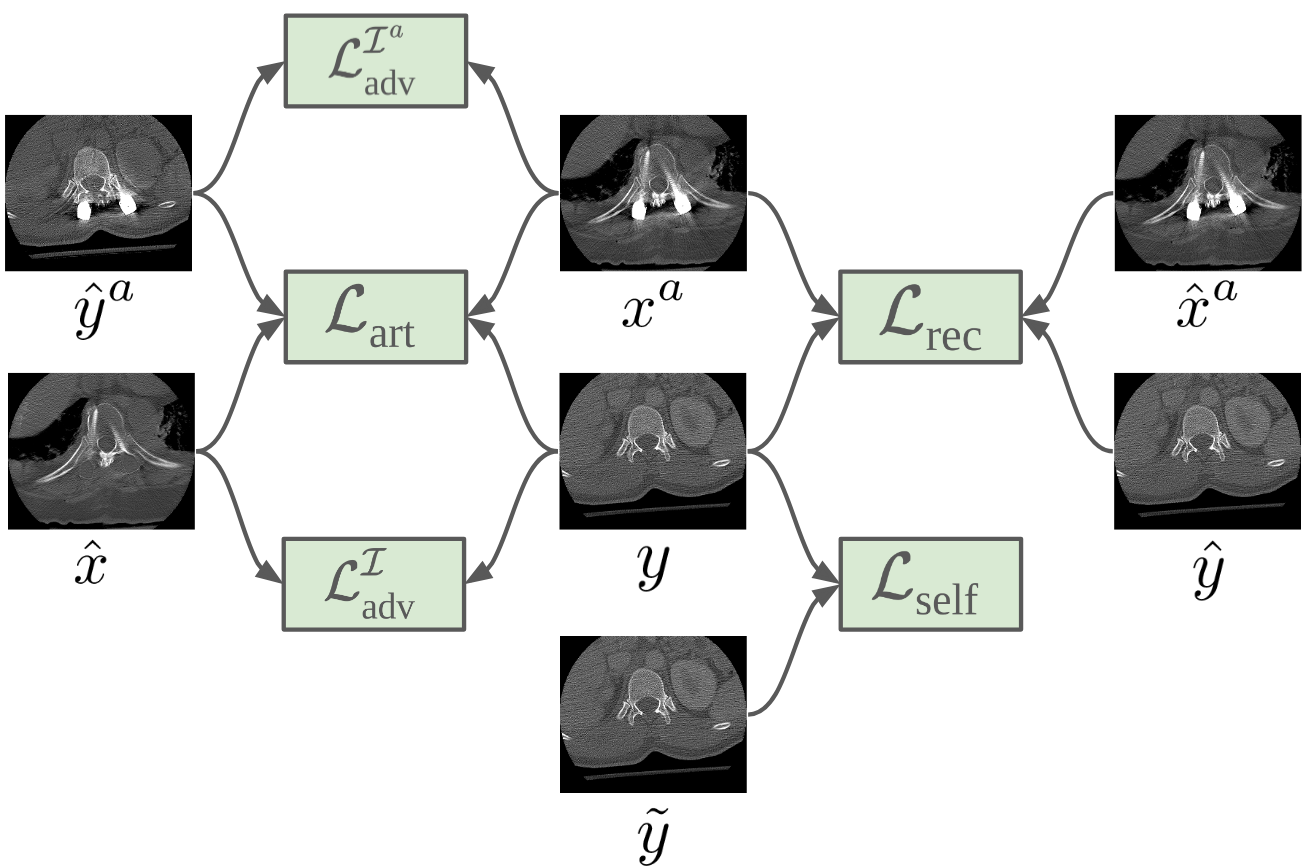}
\caption{An illustration of the relationships between the loss functions and ADN's inputs and outputs.}
\label{fig:losses}
\end{figure}

For ADN, learning an MAR model $f: \mathcal{I}^{a} \rightarrow \mathcal{I}$ means to learn the two key components $E_{\mathcal{I}^{a}}^c$ and $G_{\mathcal{I}}$. $E_{\mathcal{I}^{a}}^c$ encodes only the content of an artifact-affected image and $G_{\mathcal{I}}$ generates an artifact-free image with the encoded content code. Thus, their composition readily results in an MAR model, $f = G_{\mathcal{I}} \circ E_{\mathcal{I}^{a}}^c$. However, without paired data, it is challenging to directly address the learning of these two components. Therefore, we learn $E_{\mathcal{I}^{a}}^c$ and $G_{\mathcal{I}}$ together with other encoders and decoders in ADN. In this way, different learning signals can be leveraged to regularize the training of $E_{\mathcal{I}^{a}}^c$ and $G_{\mathcal{I}}$, and removes the requirement of paired data.

The learning aims at encouraging the outputs of the encoders and decoders to achieve the artifact disentanglement. That is, we design loss functions so that ADN outputs the intended images as denoted in Eqn.~\ref{eqn:output1}-\ref{eqn:output3}. An overview of the relationships between the loss functions and ADN's outputs is shown in Fig.~\ref{fig:losses}. We can observe that ADN enables five forms of losses, namely two adversarial losses $\mathcal{L}_{\text{adv}}^{\mathcal{I}}$ and $\mathcal{L}_{\text{adv}}^{\mathcal{I}^{a}}$, an artifact consistency loss $\mathcal{L}_{\text{art}}$, a reconstruction loss $\mathcal{L}_{\text{rec}}$ and a self-reduction loss $\mathcal{L}_{\text{self}}$.  The overall objective function is formulated as the weighted sum of these losses,
\begin{equation}
    \mathcal{L} = \lambda_{\text{adv}} (\mathcal{L}_{\text{adv}}^{\mathcal{I}} + \mathcal{L}_{\text{adv}}^{\mathcal{I}^{a}}) + \lambda_{\text{art}} \mathcal{L}_{\text{art}} + \lambda_{\text{rec}} \mathcal{L}_{\text{rec}} + \lambda_{\text{self}} \mathcal{L}_{\text{self}},
\end{equation}
where the $\lambda$'s are hyper-parameters that control the importance of each term.

\textbf{Adversarial Loss.} By manipulating the artifact component in the latent space, ADN outputs $\hat{x}$ (Eqn.~\ref{eqn:output2}) and $\hat{y}^a$ (Eqn.~\ref{eqn:output1}), where the former removes artifacts from $x^a$ and the latter adds artifacts to $y$. Learning to generate these two outputs is crucial to the success of artifact disentanglement. However, since there are no paired images, it is impossible to simply apply regression losses, such as the $L1$ or $L2$ loss, to minimize the difference between ADN's outputs and the ground truths. To address this problem, we adopt the idea of adversarial learning~\cite{gan} by introducing two discriminators $D_{\mathcal{I}^a}$ and $D_{\mathcal{I}}$ to regularize the plausibility of $\hat{x}$ and $\hat{y}^a$. On the one hand, $D_{\mathcal{I}^a}$/$D_{\mathcal{I}}$ learns to distinguish whether an image is generated by ADN or sampled from $\mathcal{I}^a$/$\mathcal{I}$. On the other hand, ADN learns to deceive $D_{\mathcal{I}^a}$ and $D_{\mathcal{I}}$ so that they cannot determine if the outputs from ADN are generated images or real images. In this way, $D_{\mathcal{I}^a}$, $D_{\mathcal{I}}$ and ADN can be trained without paired images. Formally, the adversarial loss can be written as
\begin{equation}
\begin{split}
    \mathcal{L}_{\text{adv}}^{\mathcal{I}} &= \mathbb{E}_{\mathcal{I}}[\log D_{\mathcal{I}}(y)] + \mathbb{E}_{\mathcal{I}^{a}}[1 - \log D_{\mathcal{I}}(\hat{x})]\\
    \mathcal{L}_{\text{adv}}^{\mathcal{I}^{a}} &= \mathbb{E}_{\mathcal{I}^a}[\log D_{\mathcal{I}^a}(x^a)] + \mathbb{E}_{\mathcal{I},\mathcal{I}^a}[1 - \log D_{\mathcal{I}^a}(\hat{y}^a)]\\
    \mathcal{L}_{\text{adv}} &= \mathcal{L}_{\text{adv}}^{\mathcal{I}} + \mathcal{L}_{\text{adv}}^{\mathcal{I}^{a}}
\end{split}
\end{equation}

\textbf{Reconstruction Loss.} Despite of the artifact disentanglement, there should be no information lost or model-introduced artifacts during the encoding and decoding. For artifact reduction, the content information should be fully encoded and decoded by $E_{\mathcal{I}^a}^c$ and $G_{\mathcal{I}}$. For artifact synthesis, the artifact and content components should be fully encoded and decoded by $E_{\mathcal{I}^a}^a$, $E_{\mathcal{I}}$ and $G_{\mathcal{I}^a}$. However, without paired data, the intactness of the encoding and decoding cannot be directly regularized. Therefore, we introduce two forms of reconstruction to inherently encourage the encoders and decoders to preserve the information. Specifically, ADN requires $\{E_{\mathcal{I}^a}, G_{\mathcal{I}^a}\}$ and $\{E_{\mathcal{I}}, G_{\mathcal{I}}\}$ to serve as autoencoders when encoding and decoding from the same image,
\begin{equation}
    \mathcal{L}_{\text{rec}} = \mathbb{E}_{\mathcal{I},\mathcal{I}^a}[||\hat{x}^a - x^a||_1 + ||\hat{y} - y||_1].
\end{equation}
Here, the two outputs $\hat{x}^a$ (Eqn.~\ref{eqn:output1}) and $\hat{y}$ (Eqn.~\ref{eqn:output2}) of ADN reconstruct the two inputs $x^a$ and $y$, respectively. As a common practice in image-to-image translation problem~\cite{pix2pix}, we use $L1$ loss instead of $L2$ loss to encourage sharper outputs.

\textbf{Artifact Consistency Loss.} The adversarial loss reduces metal artifacts by encouraging $\hat{x}$ to resemble a sample from $\mathcal{I}$. But the $\hat{x}$ obtained in this way is only anatomically plausible not anatomically precise, i.e., $\hat{x}$ may not be anatomically correspondent to $x^a$. A naive solution to achieve the anatomical preciseness without paired data is to directly minimize the difference between $\hat{x}$ and $x^a$ with an $L1$ or $L2$ loss. However, this will induce $\hat{x}$ to contain artifacts, and thus conflicts with the adversarial loss and compromises the overall learning. ADN addresses the anatomical preciseness by introducing an artifact consistency loss,
\begin{equation}
    \mathcal{L}_{\text{art}} = \mathbb{E}_{\mathcal{I},\mathcal{I}^a}[||(x^a - \hat{x}) - (\hat{y}^a - y)||_1].
\end{equation}
This loss is based on the observation that the difference between $x^a$ and $\hat{x}$ and the difference between $\hat{y}^a$ and $y$ should be close due to the use of the same artifact. Unlike a direct minimization of the difference between $x^a$ and $\hat{x}$,  $\mathcal{L}_{\text{art}}$ only requires $x^a$ and $\hat{x}$ to be anatomically close but not exactly close and vice versa, for $\hat{y}^a$ and $y$.

\textbf{Self-Reduction Loss.} ADN also introduces a self-reduction mechanism. It first adds artifacts to $y$ which creates $\hat{y}^a$ and then removes the artifacts from $\hat{y}^a$ which results $\tilde{y}$. Thus, we can pair $\hat{y}^a$ with $y$ to regularize the artifact reduction in Eqn.~\ref{eqn:output3} with regression,
\begin{equation}
    \mathcal{L}_{\text{self}} = \mathbb{E}_{\mathcal{I},\mathcal{I}^a}[||\tilde{y} - y||_1].
\end{equation}

\begin{figure}[t]
\centering
\includegraphics[width=\linewidth]{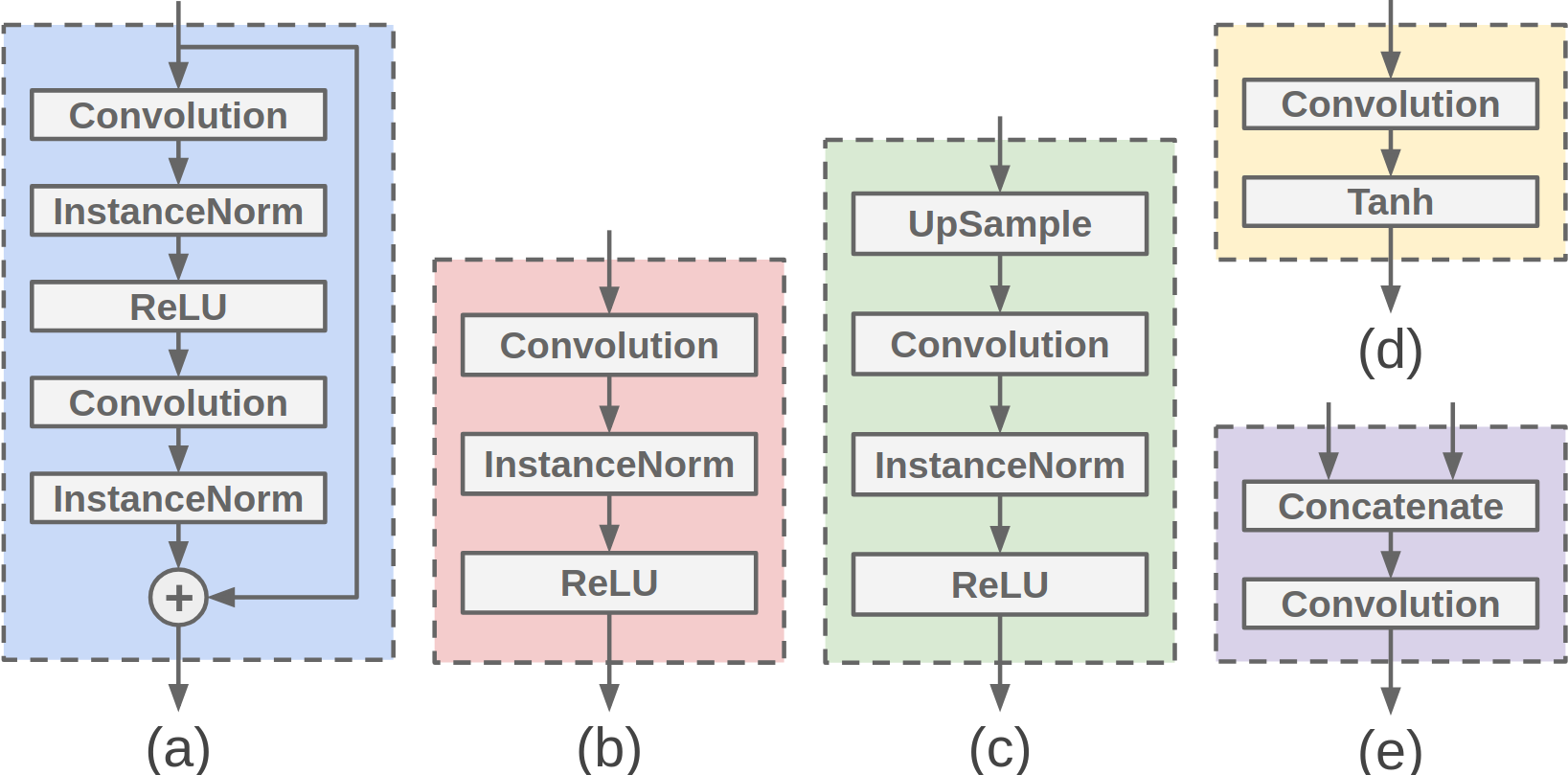}
\caption{Basic building blocks of the encoders and decoders: (a) residual block, (b) downsampling block, (c) upsampling block, (d) final block and (e) merging block.}
\label{fig:blocks}
\end{figure}

\begin{figure}[t]
\centering
\includegraphics[width=\linewidth]{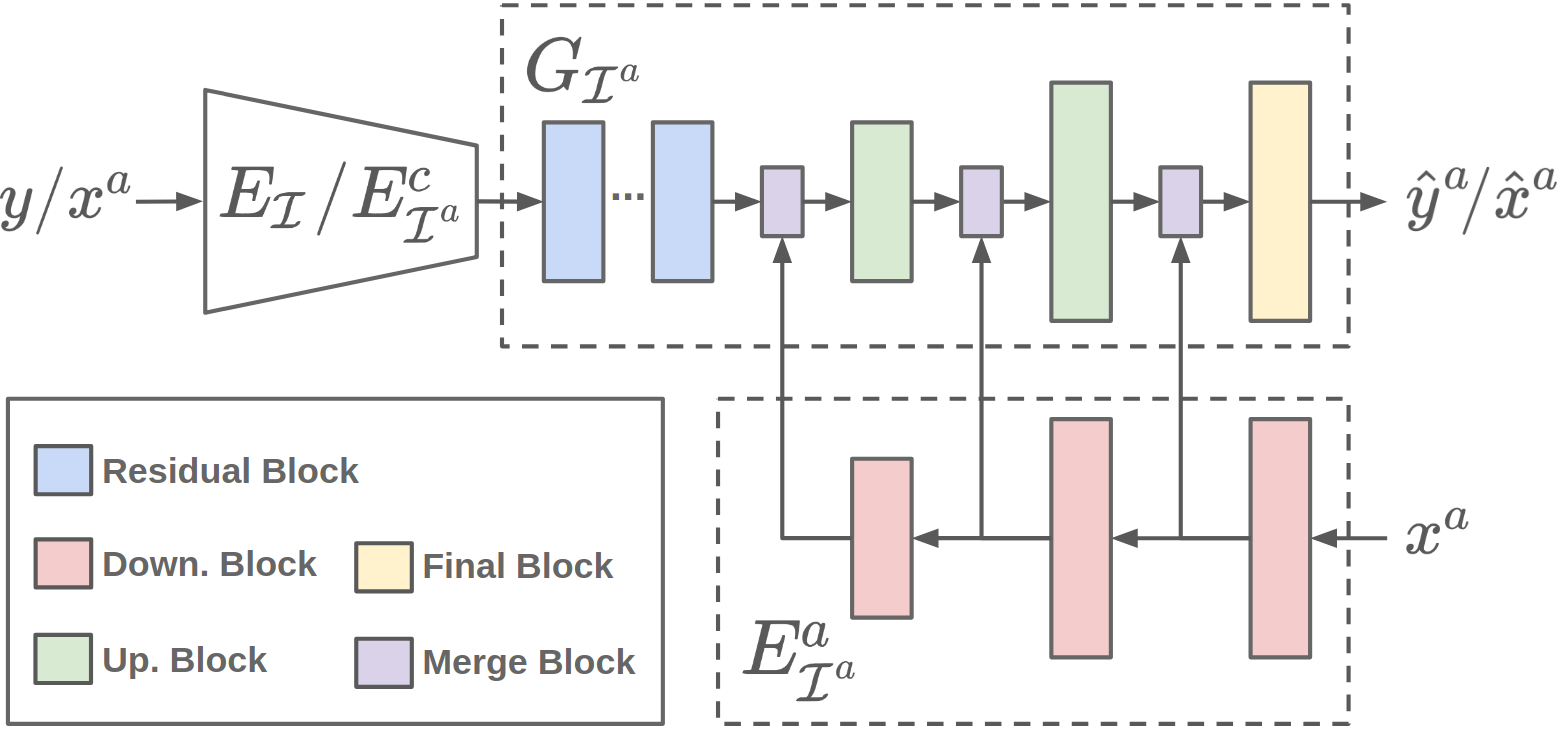}
\caption{Detailed architecture of the proposed artifact pyramid decoding (APD). The artifact-affected decoder $G_{\mathcal{I}^a}$ uses APD to effectively merge the artifact code from $E_{\mathcal{I}^a}$. }
\label{fig:apd}
\end{figure}

\subsection{Network Architectures}
\label{sec:networks}

We formulate the building components, i.e., the encoders, decoders, and discriminators, as convolutional neural networks (CNN). Table~\ref{tab:architectures} lists their detailed architectures. As we can see, the building components consist of a stack of building blocks, where some of the structures are inspired by the state-of-the-art approaches for image translation~\cite{cyclegan,munit}.

As shown in Fig.~\ref{fig:blocks}, there are five different types of blocks. The residual, downsampling and upsampling blocks are the core blocks of the encoders and decoders. The downsampling block (Fig.~\ref{fig:blocks}b) uses strided convolution to reduce the dimensionality of the feature maps for better computational efficiency. Compared with the max pooling layers, strided convolution adaptively selects the features for downsampling which demonstrates better performance for generative models~\cite{dcgan}. The residual block (Fig.~\ref{fig:blocks}a) includes residual connections to allow low-level features to be considered in the computation of high-level features. This design shows better performance for deep neural networks~\cite{resnet}. The upsampling block (Fig.~\ref{fig:blocks}c) converts feature maps back to their original dimension to generate the final outputs. We use an upsample layer (nearest neighbor interpolation) followed with a convolutional layer for the upsampling. We choose this design instead of the deconvolutional layer to avoid the ``checkerboard'' effect~\cite{checkerboard}. The padding of all the convolutional layers in the blocks of the encoders and decoders are reflection padding. It provides better results along the edges of the generated images.

It is worth noting that we propose a special way to merge the artifact code and the content code during the decoding of an artifact-affected image. We refer to this design as \textit{artifact pyramid decoding} (APD) in respect to the feature pyramid network (FPN)~\cite{fpn}. 
For artifact encoding and decoding, 
we aim to effectively recover the details of the artifacts. A feature pyramid design, which includes high-definition features with relatively cheaper costs, serves well for this purpose. Fig.~\ref{fig:apd} demonstrates the detailed architecture of APD. $E_{\mathcal{I}^a}$ consists of several downsampling blocks and outputs feature maps at different scales, i.e. a feature pyramid. $G_{\mathcal{I}^a}$ consists of a stack of residual, merge, upsample and final blocks. It generates the artifact-affected images by merging the artifact code at different scales during the decoding. The merging blocks (Fig.~\ref{fig:blocks}e) in $G_{\mathcal{I}^a}$ first concatenate the content feature maps and artifact feature maps along the channel dimension, and then use a $1\times1$ convolution to locally merge the features.

\begin{table}[]
\caption{Architecture of the building components. ``Channel (Ch.)'', ``Kernel'', ``Stride'' and ``Padding (Pad.)'' denote the configurations of the convolution layers in the blocks.}
\centering
\begin{tabular}{@{}ccccccc@{}}
\toprule
\textbf{Network}   & \textbf{Block/Layer} & \textbf{Count} & \textbf{Ch.} & \textbf{Kernel} & \textbf{Stride} & \textbf{Pad.} \\ \midrule
\multirow{4}{*}{$E_{\mathcal{I}}$ / $E_{\mathcal{I}^a}$} & down         & 1               & 64               & 7               & 1               & 3               \\
                   & down.           & 1              & 128              & 4               & 2               & 1                \\
                   & down.           & 1              & 256              & 4               & 2               & 1                \\
                   & residual        & 4              & 256              & 3               & 1               & 1                \\ \midrule
\multirow{3}{*}{$E_{a}$} & down.     & 1              & 64               & 7               & 1               & 3                \\
                   & down.           & 1              & 128              & 4               & 2               & 1                \\
                   & down.           & 1              & 256              & 4               & 2               & 1                \\ \midrule
\multirow{4}{*}{$G_{\mathcal{I}}$}   & residual       & 4                & 256             & 3               & 1                & 1                \\
                   & up.             & 1              & 128              & 5               & 1               & 2                \\
                   & up.             & 1              & 64               & 5               & 1               & 2                \\
                   & final           & 1              & 1                & 7               & 1               & 3                \\ \midrule
\multirow{7}{*}{$G_{\mathcal{I}^a}$}  & residual       & 4              & 256              & 3               & 1               & 1                \\
                   & merge           & 1              & 256              & 1               & 1               & 0                \\
                   & up.             & 1              & 128              & 5               & 1               & 2                \\
                   & merge           & 1              & 128              & 1               & 1               & 0                \\
                   & up.             & 1              & 64               & 5               & 1               & 2                \\
                   & merge           & 1              & 64               & 1               & 1               & 0                \\
                   & final           & 1              & 1                & 7               & 1               & 3                \\ \midrule
\multirow{5}{*}{$D_{\mathcal{I}}$ / $D_{\mathcal{I}^a}$} & conv          & 1               & 64              & 4                & 2               & 1             \\
                   & relu            & 1              & -                & -               & -               & -                  \\
                   & down.           & 1              & 128              & 4               & 2               & 1             \\
                   & down.           & 1              & 256              & 4               & 1               & 1             \\
                   & conv            & 1              & 1                & 4               & 1               & 1             \\ \bottomrule
\end{tabular}
\label{tab:architectures}
\end{table}

\section{Experiments}

\begin{figure}[t]
\centering
\begin{subfigure}{0.98\linewidth}
  \centering
  \includegraphics[width=\linewidth]{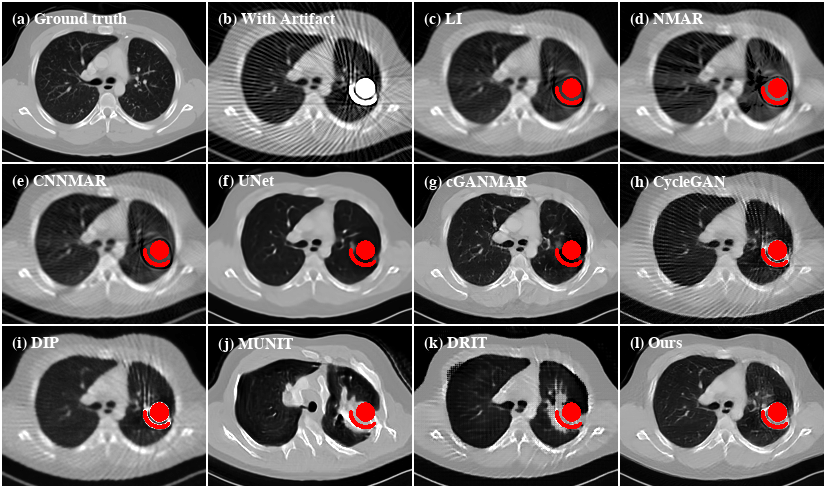}
\end{subfigure}\\
\vspace{.3em}
\begin{subfigure}{0.98\linewidth}
  \centering
  \includegraphics[width=\linewidth]{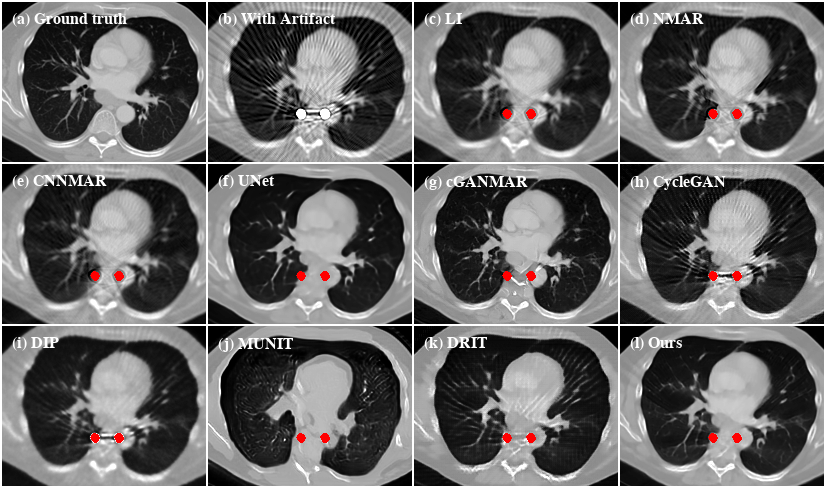}
\end{subfigure}
\caption{Qualitative comparison with baseline methods on the SYN dataset. For better visualization, we segment out the metal regions through thresholding and color them in red.}
\label{fig:syn}
\end{figure}

\subsection{Baselines}

We compare the proposed method with nine methods that are closely related to our problem. Two of the compared methods are conventional methods: LI~\cite{li} and NMAR~\cite{nmar}. These two methods are widely used approaches to MAR. 
Three of the compared methods are supervised methods: CNNMAR~\cite{cnnmartmi}, UNet~\cite{unet} and cGANMAR~\cite{cgan-mar}. CNNMAR and cGANMAR are two recent approaches that are dedicated to MAR. UNet is a general CNN framework that shows effectiveness in many image-to-image problems. The other four compared methods are unsupervised methods: CycleGAN~\cite{cyclegancvpr}, DIP~\cite{dipcvpr}, MUNIT~\cite{muniteccv} and DRIT~\cite{driteccv}. These methods are currently state-of-the-art approaches to unsupervised image-to-image translation problems.

As for the implementations of the compared methods, we use their officially released code whenever possible. For LI and NMAR, there is no official code and we adopt the implementations that are used in CNNMAR. For UNet, we use a publicly available implementation in PyTorch\footnote{github.com/milesial/Pytorch-UNet}. For cGANMAR, we train the model with the official code of Pix2Pix~\cite{pix2pix} as cGANMAR is identical to Pix2Pix at the backend.

\subsection{Datasets}

We evaluate the proposed method on one synthesized dataset and two clinical datasets. We refer to them as SYN, CL1, and CL2, respectively. For SYN, we randomly select $4,118$ artifact-free CT images from DeepLesion~\cite{deeplesion} and follow the method from CNNMAR~\cite{cnnmartmi} to synthesize metal artifacts. CNNMAR is one of the state-of-the-art supervised approaches to MAR. To generate the paired data for training, it simulates the beam hardening effect and Poisson noise during the synthesis of metal-affected polychromatic projection data from artifact-free CT images. As beam hardening effect and Poisson noise are two major causes of metal artifacts, and for a fair comparison, we apply the metal artifact synthesis method from CNNMAR in our experiments. We use $3,918$ of the synthesized pairs for training and validation and the remaining $200$ pairs for testing.

For CL1, we choose the vertebrae localization and identification dataset from Spineweb\footnote{spineweb.digitalimaginggroup.ca}. This is a challenging CT dataset for localization problems with a significant portion of its images containing metallic implants. We split the CT images from this dataset into two groups, one with artifacts and the other without artifacts. First, we identify regions with HU values greater than $2,500$ as the metal regions. Then, CT images whose largest-connected metal regions have more than 400 pixels are selected as artifact-affected images. CT images with the largest HU values less than $2,000$ are selected as artifact-free images. After this selection, the artifact-affected group contains $6,270$ images and the artifact-free group contains $21,190$ images. We withhold $200$ images from the artifact-affected group for testing.

For CL2, we investigate the performance of the proposed method under a more challenging {\it cross-modality} setting. Specifically, the artifact-affected images of CL2 are from a cone-beam CT (CBCT) dataset collected during spinal interventions. Images from this dataset are very noisy. The majority of them contain metal artifacts while the metal implants are mostly not within the imaging field of view. There are in total $2,560$ CBCT images from this dataset, among which 200 images are withheld for testing. For the artifact-free images, we reuse the CT images collected from CL1.

Note that LI, NMAR, and CNNMAR require the availability of raw X-ray projections which however are not provided by SYN, CL1, and CL2. Therefore, we follow the literature~\cite{cnnmartmi} by synthesizing the X-ray projections via forward projection. For SYN, we first forward project the artifact-free CT images and then mask out the metal traces. For CL1 and CL2, there are no ground truth artifact-free CT images available. Therefore, the X-ray projections are obtained by forward projecting the artifact-affected CT images. The metal traces are also segmented and masked out for projection interpolation.

\subsection{Training and testing}

We implement our method under the PyTorch deep learning framework\footnote{pytorch.org} and use the Adam optimizer with $\num{1e-4}$ learning rate to minimize the objective function. For the hyper-parameters, we use $\lambda_{\text{adv}}^{\mathcal{I}}=\lambda_{\text{adv}}^{\mathcal{I}^a}=1.0$, $\lambda_{\text{rec}}=\lambda_{\text{self}}=\lambda_{\text{art}}=20.0$ for SYN and CL1, and use $\lambda_{\text{adv}}^{\mathcal{I}}=\lambda_{\text{adv}}^{\mathcal{I}^a}=1.0$, $\lambda_{\text{rec}}=\lambda_{\text{self}}=\lambda_{\text{art}}=5.0$ for CL2.

Due to the artifact synthesis, SYN contains paired images for supervised learning. To simulate the unsupervised setting for SYN, we evenly divide the $3,918$ training pairs into two groups. For one group, only artifact-affected images are used and their corresponding artifact-free images are withheld. For the other group, only artifact-free images are used and their corresponding artifact-affected images are withheld. During the training of the unsupervised methods, we randomly select one image from each of the two groups as the input.

To train the supervised methods with CL1, we first synthesize metal artifacts using the images from the artifact-free group of CL1. Then, we train the supervised methods with the synthesized pairs. During testing, the trained models are applied to the testing set containing only clinical metal artifact images. To train the unsupervised methods, we randomly select one image from the artifact-affected group and the other from the artifact-free group as the input. In this way, the artifact-affected images and artifact-free images are sampled evenly during training which helps with the data imbalance between the artifact-affected and artifact-free groups.

For CL2, synthesizing metal artifacts is not possible due to the unavailability of artifact-free CBCT images. Therefore, for the supervised methods, we directly use the models trained for CL1. In other words, the supervised methods are trained on synthesized CT images (from CL1) and tested on clinical CBCT images (from CL2). For the unsupervised models, each time we randomly select one artifact-affected CBCT image and one artifact-free CT image as the input for training. 

\subsection{Performance on synthesized data}

\begin{table}[]
\centering
\caption{Quantitative comparison with baseline methods on the SYN dataset.}
\label{tab:metrics}
\resizebox{0.75\columnwidth}{!}{
\begin{tabular}{@{}llcc@{}}
\toprule
\multirow{2}{*}{}            & \multicolumn{1}{c}{\multirow{2}{*}{\textbf{Method}}} & \multicolumn{2}{c}{\textbf{\hspace{.5em} \vspace{.2em}  Metrics}} \\
                             & \multicolumn{1}{c}{}                                 & \textbf{PSNR}     & \textbf{SSIM}    \\ \midrule
\multirow{2}{*}{Conventional} & LI~\cite{li}                                         & 32.0              & 91.0             \\
                             & NMAR~\cite{nmar}                                     & 32.1              & 91.2             \\ \midrule
\multirow{3}{*}{Supervised}  & CNNMAR~\cite{cnnmartmi}                              & 32.5              & 91.4             \\
                             & UNet~\cite{unet}                                     & \textbf{34.8}     & 93.1             \\
                             & cGANMAR~\cite{cgan-mar}                              & 34.1              & \textbf{93.4}    \\ \midrule
\multirow{4}{*}{Unsupervised}    & CycleGAN~\cite{cyclegan}                             & 30.8              & 72.9             \\
                             & DIP~\cite{dip}                                       & 26.4              & 75.9             \\
                             & MUNIT~\cite{munit}                                   & 14.9              & 7.5              \\
                             & DRIT~\cite{drit}                                     & 25.6              & 79.7             \\
                             & Ours                                                  & \underline{33.6}  & \underline{92.4} \\ \bottomrule
\end{tabular}}
\end{table}

\begin{figure}[t]
\centering
\begin{subfigure}{0.98\linewidth}
  \centering
  \includegraphics[width=\linewidth]{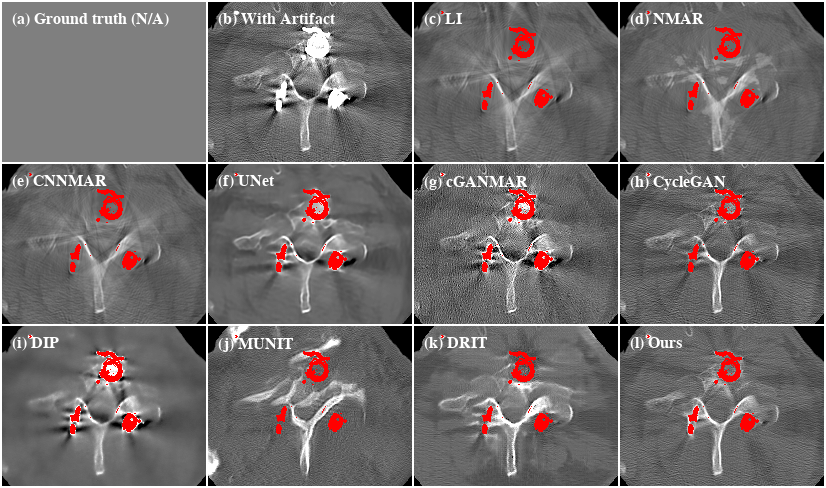}
\end{subfigure}\\
\vspace{.3em}
\begin{subfigure}{0.98\linewidth}
  \centering
  \includegraphics[width=\linewidth]{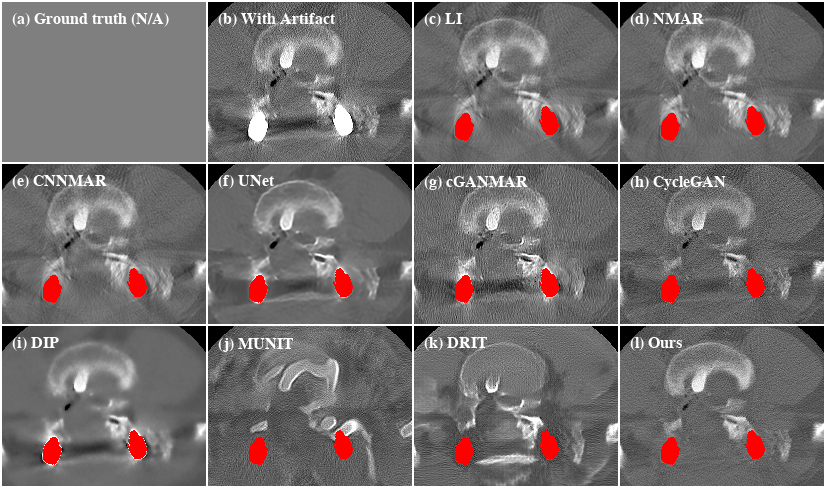}
\end{subfigure}
\caption{Qualitative comparison with baseline methods on the CL1 dataset. For better visualization, we obtain the metal regions through thresholding and color them with red.}
\label{fig:cl1}
\end{figure}

SYN contains paired data, allowing for both quantitative and qualitative evaluations. Following the convention in the literature, we use peak signal-to-noise ratio (PSNR) and structural similarity index (SSIM) as the metrics for the quantitative evaluation. For both metrics, the higher values are better. Table~\ref{tab:metrics} and Fig.~\ref{fig:syn} show the quantitative and qualitative evaluation results, respectively.

We observe that our proposed method performs significantly better than the other unsupervised methods. MUNIT focuses more on diverse and realistic outputs (Fig.~\ref{fig:syn}j) with less constraint on structural similarity. CycleGAN and DRIT perform better as both the two models also require the artifact-corrected outputs to be able to transform back to the original artifact-affected images. Although this helps preserve content information, it also encourages the models to keep the artifacts. Therefore, as shown in Fig.~\ref{fig:syn}h and \ref{fig:syn}k, the artifacts cannot be effectively reduced. DIP does not reduce much metal artifact in the input image (Fig.~\ref{fig:syn}i) as it is not designed to handle the more structured metal artifact.

We also find that the performance of our method is on par with the conventional and supervised methods. The performance of UNet is close to that of cGANMAR which at its backend uses an UNet-like architecture. However, due to the use of GAN, cGANMAR produces sharper outputs (Fig.~\ref{fig:syn}g) than UNet (Fig.~\ref{fig:syn}f). As for PSNR and SSIM, both methods only slightly outperform our method. LI, NMAR, and CNNMAR are all projection interpolation based methods. NMAR is better than LI as it uses prior images to guide the projection interpolation. CNNMAR uses CNN to learn the generation of the prior images and thus shows better performance than NMAR. As we can see, ADN performs better than these projection interpolation based approaches both quantitatively and qualitatively.

\subsection{Performance on clinical data}

\begin{figure}[t]
\centering
\begin{subfigure}{0.98\linewidth}
  \centering
  \includegraphics[width=\linewidth]{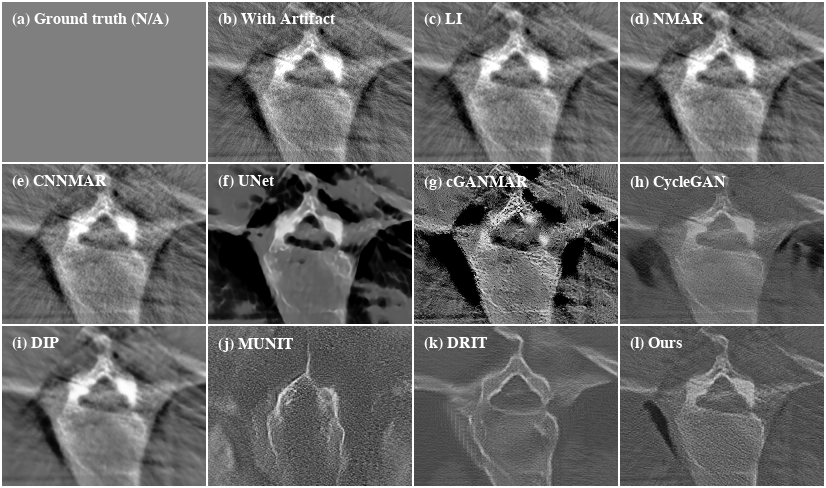}
\end{subfigure}\\
\vspace{.3em}
\begin{subfigure}{0.98\linewidth}
  \centering
  \includegraphics[width=\linewidth]{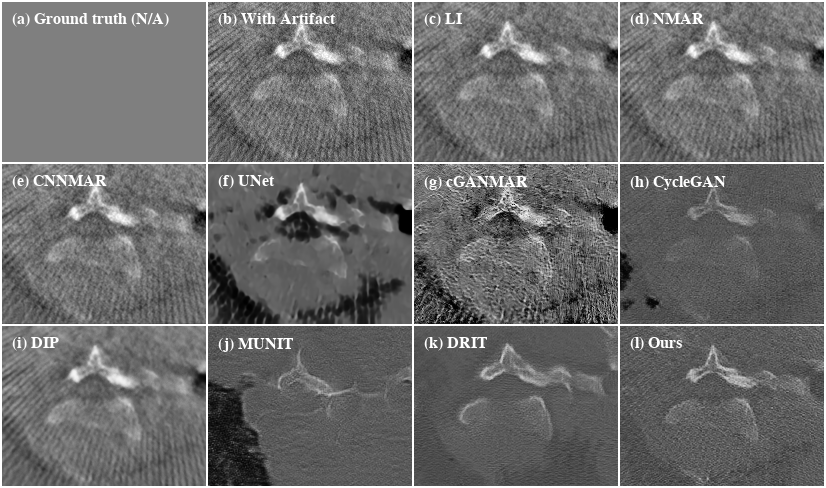}
\end{subfigure}
\caption{Qualitative comparison with baseline methods on the CL2 dataset.}
\label{fig:cl2}
\end{figure}

Next, we investigate the performance of the proposed method on clinical data. Since there are no ground truths available for the clinical images, only qualitative comparisons are performed. The qualitative evaluation results of CL1 are shown in Fig.~\ref{fig:cl1}. Here, all the supervised methods are trained with paired images that are synthesized from the artifact-free group of CL1. We can see that UNet and cGANMAR do not generalize well when applied to clinical images (Fig.~\ref{fig:cl1}f and \ref{fig:cl1}g). LI, NMAR, and CNNMAR are more robust as they correct the artifacts in the projection domain. However, the projection domain corrections also introduce secondary artifacts (Fig.~\ref{fig:cl1}c, \ref{fig:cl1}d and \ref{fig:cl1}e). For the more challenging CL2 dataset (Fig.~\ref{fig:cl2}), all the supervised methods fail. This is not totally unexpected as the supervised methods are trained using only CT images because of the lack of artifact-free CBCT images. As the metallic implants of CL2 are not within the imaging field of view, there are no metal traces available and the projection interpolation based methods do not work (Fig.~\ref{fig:cl2}c, \ref{fig:cl2}d and \ref{fig:cl2}e). Similar to the cases with SYN, the other unsupervised methods also show inferior performances when evaluated on both the CL1 and CL2 datasets. In contrast, our method removes the dark shadings and streaks significantly without introducing secondary artifacts.

\subsection{Ablation study}

\begin{figure}[t]
\centering
\begin{subfigure}{0.98\linewidth}
  \centering
  \includegraphics[width=\linewidth]{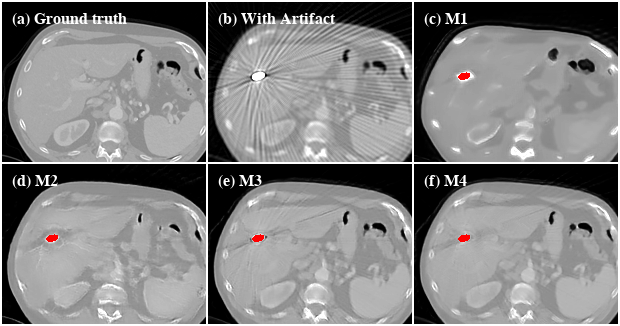}
\end{subfigure}\\
\vspace{.3em}
\begin{subfigure}{0.98\linewidth}
  \centering
  \includegraphics[width=\linewidth]{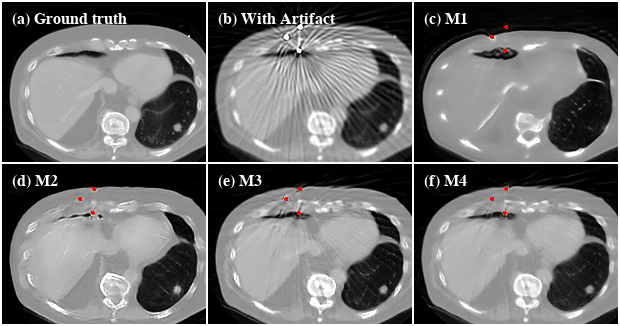}
\end{subfigure}
\caption{Qualitative comparison of different variants of ADN. The compared models (M1-M4) are trained with different combinations of the loss functions discussed in Sec.~\ref{sec:learn}.}
\label{fig:ablation}
\end{figure}

We perform an ablation study to understand the effectiveness of several designs of ADN. All the experiments are conducted with the SYN dataset so that both the quantitative and qualitative performances can be analyzed. Table~\ref{tab:ablation} and Fig.~\ref{fig:ablation} show the experimental results, where the performances of ADN (M4) and its three variants (M1-M3) are compared. M1 refers to the model trained with only the adversarial loss $\mathcal{L_{\text{adv}}}$. M2 refers to the model trained with both the adversarial loss $\mathcal{L_{\text{adv}}}$ and the reconstruction loss $\mathcal{L_{\text{rec}}}$. M3 refers to the model trained with the adversarial loss $\mathcal{L_{\text{adv}}}$, the reconstruction loss $\mathcal{L_{\text{rec}}}$, and the artifact consistency loss $\mathcal{L_{\text{art}}}$. M4 refers to the model trained with all the losses, i.e., ADN. We use M4 and ADN interchangeably in the experiments.

From Fig.~\ref{fig:ablation}, we can observe that M1 generates artifact-free images that are structurally similar to the inputs. However, with only adversarial loss, there is no support that the content of the generated images should exactly match the inputs. Thus, we can see that many details of the inputs are lost and some anatomical structures are mismatched. In contrast, the results from M2 maintain most of the anatomical details of the inputs. This demonstrates that learning to reconstruct the inputs is helpful to guide the model to preserve the details of the inputs. However, as the reconstruction loss is applied in a self-reconstruction manner, there is no direct penalty for the anatomical reconstruction error during the artifact reduction. Thus, we can still observe some minor anatomical imperfections from the outputs of M2.

M3 improves M2 by including the artifact consistency loss. This loss directly measures the pixel-wise anatomical differences between the inputs and the generated outputs. As shown in Fig.~\ref{fig:ablation}, the results of M3 precisely preserve the content of inputs and suppress most of the metal artifacts. For M4, we can find that the outputs are further improved. This shows that the self-reduction mechanism, which allows the model to reduce synthesized artifacts, is indeed helpful. The quantitative results are provided in Table~\ref{tab:ablation}. We can see that they are consistent with our qualitative observations in Fig.~\ref{fig:ablation}.

\begin{table}[]
\centering
\caption{Quantitative comparison of different variants of ADN. The compared models (M1-M4) are trained with different combinations of the loss functions discussed in Sec.~\ref{sec:learn}.}
\label{tab:ablation}
\resizebox{0.6\columnwidth}{!}{
\begin{tabular}{@{}lcc@{}}
\toprule
\multicolumn{1}{c}{\multirow{2}{*}{\textbf{Method}}} & \multicolumn{2}{c}{\hspace{.5em} \vspace{.2em} \textbf{Metrics}} \\ 
\multicolumn{1}{c}{}                                 & \textbf{PSNR}     & \textbf{SSIM}    \\ \midrule
M1 ($\mathcal{L}_\text{adv}$ only)                     & 21.7              & 61.5             \\
M2 (M1 with $\mathcal{L}_\text{rec}$)                  & 26.3              & 82.1             \\
M3 (M2 with $\mathcal{L}_\text{art}$)                  & 32.8              & 91.6             \\
M4 (M3 with $\mathcal{L}_\text{self}$)                 & \bf{33.6}         & \bf{92.4}        \\ \bottomrule
\end{tabular}}
\end{table}

\subsection{Artifact Synthesis}

\begin{figure}
\centering
\includegraphics[width=\linewidth]{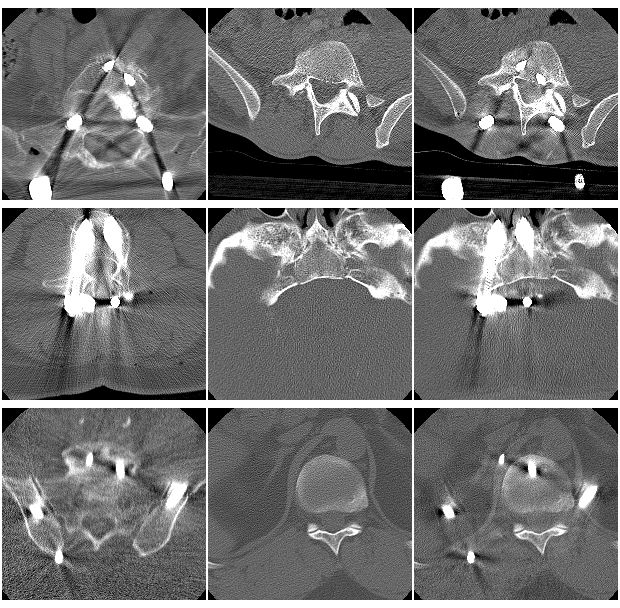}
\caption{Metal artifact transfer. Left: the clinical images with metal artifacts $x^a$. Middle: the clinical images without metal artifacts $y$. Right: the metal artifacts on the left column transferred to the artifact-free images in the middle $\hat{y}^a$.}
\label{fig:art_syn}
\end{figure}

In addition to artifact reduction, ADN also supports unsupervised artifact synthesis. This functionality arises from two designs. First, the adversarial loss $\mathcal{L}^{\mathcal{I}^a}_{adv}$ encourages the output $\hat{y}^a$ to be a sample from $\mathcal{I}^a$, i.e. the metal artifact should look real. Second, the artifact consistency loss $\mathcal{L}_{art}$ induces $\hat{y}^a$ to contain the metal artifacts from $x^a$ and suppresses the synthesis of the content component from $x$. This section investigates the effectiveness of these two designs.  The experiments are performed with the CL1 dataset as learning to synthesize clinical artifacts is more practical and challenging than learning to synthesize the artifacts from SYN, whose artifacts are already synthesized. Fig.~\ref{fig:art_syn} shows the experimental results, where each row is an example of artifact synthesis. Images on the left are the clinical images with metal artifacts. Images in the middle are the clinical images without artifacts. Images on the right are the artifact synthesis results by transferring the artifacts from the left image to the middle image. As we can see, except the positioning of the metal implants, the synthesized artifacts look realistic. The metal artifacts merge naturally into the artifact-free images making it really challenging to notice that the artifacts are actually synthesized. More importantly, it is only the artifacts that are transferred and almost no content is transferred to the artifact-free images. Note that our model is data-driven. If there is an anatomical structure or lesion that looks like metal artifacts, it might also be transferred.

\section{Discussions}

\textbf{Applications to Artifact Reduction.} Given the flexibility of ADN, we expect many applications to artifact reduction in medicine, where obtaining paired data is usually impractical. First, as we have already demonstrated, ADN can be applied to address metal artifacts. It reduces metal artifacts directly with CT images, which is critical to the scenarios when researchers or healthcare practitioners have no access to the raw projection data as well as the associated reconstruction algorithms. For the manufacturers, ADN can be applied in a post-processing step to further improve the in-house MAR algorithm that addresses metal artifacts in the projection data during the CT reconstruction.

Second, even though our problem under investigation is MAR, ADN should work with other artifact reduction problems as well. In the problem formulation, ADN does not make any assumption about the nature of the artifacts. Therefore, if we change to other artifact reduction problems such as deblurring, destreaking, denoising, etc., ADN should also work. Actually, in the experiments, the input images from CL1 (Fig.~\ref{fig:cl1}b) are slightly noisy while the outputs of ADN are more smooth. Similarly, input images from CL2 (Fig.~\ref{fig:cl2}b) contain different types of artifacts, such as noise, streaking artifacts and so on, and ADN handles them well.

\textbf{Applications to Artifact Synthesis.} By combining $E_{\mathcal{I}^{a}}^a$, $E_{\mathcal{I}}$ and $G_{\mathcal{I}^a}$, ADN can be applied to synthesize artifacts in an artifact-free image. As we have shown in Fig.~\ref{fig:art_syn}, the synthesized artifacts look natural and realistic, which may potentially have practical applications in medical image analysis. For example, a CT image segmentation model may not work well when metal artifacts are present as there are not enough metal-affected images in the dataset. By using ADN, we could significantly increase the number of metal-affected images in the dataset via the realistic metal artifact synthesis. In this way, ADN may potentially improve the performance of the CT segmentation model.

\section{Conclusions}

We present an unsupervised learning approach to MAR. Through the development of an artifact disentanglement network, we have shown how to leverage artifact disentanglement to achieve different forms of image translations as well as self-reconstructions that eliminate the requirement of paired images for training. To understand the effectiveness of this approach, we have performed extensive evaluations on one synthesized and two clinical datasets. The evaluation results demonstrate the feasibility of using unsupervised learning method to achieve comparable performance to the supervised methods with synthesized dataset. More importantly, the results also show that directly learning MAR from clinical CT images under an unsupervised setting is a more feasible and robust approach than simply applying the knowledge learned from synthesized data to clinical data. We believe our findings in this work will stimulate more applicable research for medical image artifact reduction under an unsupervised setting.

\appendices
\section*{Acknowledgment}

This work was supported in part by NSF award \#1722847 and the Morris K. Udall Center of Excellence in Parkinson's Disease Research by NIH.

\ifCLASSOPTIONcaptionsoff
  \newpage
\fi



\bibliographystyle{IEEEtran}
\bibliography{IEEEabrv,references}

\end{document}